\documentclass[aps, twocolumn, pra, showpacs, superscriptaddress]{revtex4}
\usepackage{graphicx}
\usepackage{dcolumn}
\usepackage{bm}
\usepackage{amsmath}

\begin{document}

\title{High-order fidelity and quantum phase transition for the Heisenberg
chain with next-nearest-neighbor interaction}
\author{Li Wang}
\affiliation{Institute of Physics, Chinese Academy of Sciences, Beijing 100080, China}
\author{Shi-Jian Gu}
\affiliation{Department of Physics and ITP, The Chinese University of Hong Kong, Hong
Kong, China}
\author{Shu Chen}\email{schen@aphy.iphy.ac.cn}
\affiliation{Institute of Physics, Chinese Academy of Sciences, Beijing 100080,
China}
\date{\today }

\begin{abstract}
In this article, we study the high order term of the fidelity of the
Heisenberg chain with next-nearest-neighbor interaction and analyze
its connection with quantum phase transition of
Beresinskii-Kosterlitz-Thouless type happened in the system. We
calculate the fidelity susceptibility of the system and find that
although the phase transition point can't be well characterized by
the fidelity susceptibility, it can be effectively picked out by the
higher order of the ground-state fidelity for finite-size systems.
\end{abstract}

 \maketitle

\section{Introduction}

Quantum phase transitions (QPTs) of a quantum many-body system have been
attracting the persistent interest of physical researchers in recent years.
Due to the diversity of quantum phases and QPTs, finding universal ways or
methods to characterize QPTs is very meaningful and urgent. From the
viewpoint of Landau-Ginzburg theory which has been widely accepted and known
in condensed matter physics \cite{sachdev}, QPT is connected with the
corresponding order parameter and symmetry breaking. However, there are also
some QPTs which cannot be well understood under the Landau-Ginzburg
paradigm, such as the topological phase transitions \cite{xgwen} and
Beresinskii-Kosterlitz-Thouless (BKT) phase transitions \cite%
{beresinskii,kosterlitz}. Recently, an increasing research effort has been
focused on the role of ground-state fidelity in characterizing QPTs\cite%
{Gu_review,htquan,zanardi06,
hqzhou,YouWL07,zanardi07PRl,schen07pre,schen08pra,buonsante,mfyang,ZhouPRL}.
As a basic concept in quantum information science, the fidelity
measures the similarity between two states and is simply defined as
modulus of their overlap \cite{zanardi06}. The fidelity approach
provides us a novel way to understand QPTs from the viewpoint of
quantum information theory. So far QPTs in various quantum many-body
systems \cite{YouWL07,hqzhou,mfyang,
qhchen,mfyang08,WangXG08,schen07pre,schen08pra,Paunkovic,Venuti,buonsante,AHamma07,
abasto,YangS,zanardi07PRl,WangXG,Zhou09,Zhou0803,ZhouPRL} have been
shown to be well characterized by the ground-state fidelity or
fidelity susceptibility which is the leading term of the fidelity \cite%
{YouWL07,zanardi07PRl}.

Generally, one may expect that the structure of the ground states at
the different phases is basically different and should reveal itself
by some sort of singular behavior in the ground state fidelity or
the fidelity susceptibility at the transition point \cite{zanardi06,
hqzhou}. Despite its great successes of application in various
systems, this intuitive idea turns out to be not complete
\cite{schen07pre, mfyang, schen08pra, YouWL07,mfyang08}. Although
the fidelity and the fidelity susceptibility can be used to describe
first- and second-order QPTs\cite{schen08pra}, as well as the
topological QPTs \cite{AHamma07,abasto,YangS,Zhou0803} successfully,
nevertheless there are also some ambiguous cases for that both the
two methods mentioned above do not work very effectively
\cite{schen07pre, schen08pra,YouWL07,mfyang08}. Very recently, the
controversial issue of BKT phase transition and ground state
fidelity has been studied in Ref. \cite{Zhou09} from a perspective
of matrix product states which essentially depend on a classical
simulations of quantum lattice systems \cite{ZhouPRL}.

In case that the leading term of the fidelity (fidelity
susceptibility) works not very effectively, the higher order term in
the fidelity may be worth studying. Up to now, there is still lack
of literature concerning this part of the fidelity. Here, in this
paper, we make an attempt on investigating the effect of higher
order term of the fidelity on the characterization of the BKT-type
phase transition happened in the Heisenberg chain with
next-nearest-neighbor (NNN) interaction \cite{Haldane}. We will show
that although the fidelity and fidelity susceptibility cannot
effectively characterize the BKT-type phase transition point for the
Heisenberg chain with NNN interaction, the higher order term of the
fidelity gives a good attempt on detecting such a transition.

Our paper is organized as follows. In Sec. II, we display the formulism of
the higher order term of the fidelity. The subsequent section is devoted to
the calculation of the higher order term of the fidelity for the model of
Heisenberg chain with NNN interaction and show its connection to the quantum
phase transition of the system. A brief summary is given in Sec. \ref%
{sec:sum}.

\section{Higher order of the fidelity}

\label{sec:highorder}

As usual, the ground state fidelity is defined as the modulus of the overlap
between $|\Psi_0(\lambda) \rangle$ and $| \Psi_0(\lambda+\delta\lambda)
\rangle$, i.e.
\begin{equation}
F(\lambda, \delta\lambda) =\left| f(\lambda, \lambda+\delta\lambda) \right |
= \left | \langle \Psi_0(\lambda)| \Psi_0(\lambda+\delta\lambda) \rangle
\right | ,  \tag{1}  \label{eqF}
\end{equation}
where $\Psi_0(\lambda)$ is the ground-state wavefunction of Hamiltonian $%
H=H_0 + \lambda H_I$, $\lambda$ is the driving parameter and $\delta \lambda$
is a small deviation in the parameter space of $\lambda$. The fidelity
susceptibility denotes only the leading term of the fidelity.
Straightforwardly, one can get the higher order term of the fidelity
following similar expansion in deriving the fidelity susceptibility \cite%
{YouWL07}. By using the Taylor expansion, the overlap between two
wavefunction $|\Psi _{0}(\lambda )\rangle $ and $|\Psi _{0}(\lambda +\delta
\lambda )\rangle $ can be expanded to an arbitrary order of $\delta\lambda$,
i.e.
\begin{equation*}
f(\lambda ,\lambda +\delta \lambda )=1+\sum_{n=1}^{\infty }\frac{(\delta
\lambda )^{n}}{n!}\left\langle \Psi _{0}(\lambda )\left\vert \frac{\partial
^{n}}{\partial \lambda ^{n}}\Psi _{0}(\lambda )\right. \right\rangle .
\tag{2}  \label{eqf}
\end{equation*}
Therefore, the fidelity becomes
\begin{align*}
F^{2}=&1+\sum_{n=1}^{\infty }\frac{(\delta \lambda )^{n}}{n!}\left\langle
\Psi _{0}\left\vert \frac{\partial ^{n}}{\partial \lambda ^{n}}\Psi
_{0}\right. \right\rangle+  \notag \\
&\sum_{n=1}^{\infty }\frac{(\delta \lambda )^{n}}{n!}\left\langle \left.
\frac{\partial ^{n}}{\partial \lambda ^{n}}\Psi _{0}\right\vert \Psi
_{0}\right\rangle +  \notag \\
&\sum_{m,n=1}^{\infty }\frac{(\delta \lambda )^{m+n}}{m!n!}\left\langle \Psi
_{0}\left\vert \frac{\partial ^{n}}{\partial \lambda ^{n}}\Psi _{0}\right.
\right\rangle \left\langle \left. \frac{\partial ^{m}}{\partial \lambda ^{m}}%
\Psi _{0}\right\vert \Psi _{0}\right\rangle.  \tag{3}  \label{eqF2}
\end{align*}
We note that $\frac {\partial^{n}} {\partial \lambda^{n}} \langle
\Psi_0(\lambda)| \Psi_0(\lambda) \rangle =0$ and use the relation for a
given $n$
\begin{equation*}
\sum_{m=0}^{n}\frac{n!}{m!(n-m)!}\left\langle \left. \frac{\partial ^{m}}{%
\partial \lambda ^{m}}\Psi _{0}\right\vert \frac{\partial ^{n-m}}{\partial
\lambda ^{n-m}}\Psi _{0}\right\rangle =0 ,  \tag{4}  \label{relation}
\end{equation*}
then we can simplify the expression of (\ref{eqF2}) into
\begin{equation}
F^{2}=1-\sum_{l=1}^{\infty }(\delta \lambda )^{l}\chi _{F}^{(l)}  \tag{5}
\label{F2simple}
\end{equation}%
where
\begin{equation}
\chi _{F}^{(l)}=\sum_{l=m+n}\frac{1}{m!n!}\left\langle \left. \frac{\partial
^{m}}{\partial \lambda ^{m}}\Psi _{0}\right\vert \hat {P} \left\vert \frac{%
\partial ^{n}}{\partial \lambda ^{n}}\Psi _{0}\right. \right\rangle ,
\tag{6}  \label{eq:higherorderdiff}
\end{equation}
with the projection operator $\hat {P}$ defined as $\hat {P}=1- |\Psi_0
\rangle \langle \Psi_0 |$. It is easy to check that $\chi _{F}^{(1)}$ is
zero and $\chi _{F}^{(2)}$ the fidelity susceptibility \cite{YouWL07}.

Next we shall consider the third order fidelity $\chi _{F}^{(3)}$ and apply
it to judge the phase transition in the spin chain model with NNN exchanges.
Alternatively, one can directly derive the expression of $\chi _{F}^{(3)}$
from the perturbation expansion of the GS wavefunction. According the
perturbation theory, the GS wavefunction, up to the second order, is
\begin{align*}
|\Psi _{0}(\lambda +\delta \lambda )\rangle =& |\Psi _{0}\rangle +\delta
\lambda \sum_{n\neq 0}\frac{H_{I}^{n0}|\Psi _{n}\rangle }{E_{0}-E_{n}} \\
& +\left( \delta \lambda \right) ^{2}\sum_{m,n\neq 0}\frac{%
H_{I}^{nm}H_{I}^{m0}|\Psi _{n}\rangle }{(E_{0}-E_{m})(E_{0}-E_{n})} \\
& -\left( \delta \lambda \right) ^{2}\sum_{n\neq 0}\frac{%
H_{I}^{00}H_{I}^{n0}|\Psi _{n}\rangle }{(E_{0}-E_{n})^{2}} \\
& -\frac{\left( \delta \lambda \right) ^{2}}{2}\sum_{n\neq 0}\frac{%
H_{I}^{0n}H_{I}^{n0}|\Psi _{0}\rangle }{(E_{0}-E_{n})^{2}}.
\end{align*}%
The 3rd order term $\chi _{F}^{(3)}$, which is proportional to the
3rd order derivative of GS fidelity, can be then directly extracted
from eq. (\ref{F2simple}):
\begin{equation}
\chi _{F}^{(3)}=\sum_{m,n\neq 0}\frac{2H_{I}^{0m}H_{I}^{mn}H_{I}^{n0}}{%
(E_{0}-E_{m})(E_{0}-E_{n})^{2}}-\sum_{n\neq 0}\frac{2H_{I}^{00}\left\vert
H_{I}^{n0}\right\vert ^{2}}{(E_{0}-E_{n})^{3}}.  \tag{8}
\label{eq:higheroderperturb}
\end{equation}

Eqs. (\ref{eq:higherorderdiff}) and (\ref{eq:higheroderperturb}) present the
main formulism of the higher order expansion of the fidelity. So far the
explicit physical meaning of the high order term in the fidelity is still
not clear. The expression of 3rd fidelity bears the similarity to its
correspondence of the 3rd derivative of GS energy which has the following
form
\begin{equation}
\frac {\partial^3 E} {\partial \lambda^3} = \sum_{m,n\neq0}\frac{6
H_I^{0n}H_I^{nm}H_I^{m0}}{(E_0-E_m)(E_0-E_n)} -\sum_{n\neq0}\frac{6
H_I^{00}\left|H_I^{n0}\right|^2}{(E_0-E_n)^2}.   \tag{9}
\end{equation}
Obviously, the 3rd fidelity is more divergent than the 3rd
derivative of GS energy. Similar connection between the fidelity
susceptibility and 2nd derivative of GS energy has been unveiled
\cite{schen08pra}. Generally the $n $-th order fidelity is much more
divergent than its counterpart of $n$-th order derivative of GS
energy, therefore an $n$-th order QPT can be certainly detected by
the $n$-th order fidelity. However, this conclusion does not exclude
the possibility that $n$-th order fidelity can detect a even higher
order or infinite order QPT. A concrete example has been given in Ref. \cite%
{mfyang08}, where a QPT of higher than second order was singled out
unambiguously by using the fidelity susceptibility despite the corresponding
second derivative of the ground-state energy density showing no signal of
divergence. So far no example of BKT-type QPT unambiguously detected by
fidelity susceptibility has been given. Next we shall attempt to apply the
third-order fidelity to study the BKT-type transition in a the spin chain
model with NNN exchanges.

\section{The model and the calculation of 3rd order fidelity}

\label{sec:model}

Now we turn to the one-dimensional Heisenberg chain with the NNN coupling
described by the Hamiltonian
\begin{equation}
H(\lambda )=\sum_{j=1}^{L}\left( \hat{s}_{j}\hat{s}_{j+1}+\lambda \hat{s}_{j}%
\hat{s}_{j+2}\right) ,  \tag{10}  \label{Ham}
\end{equation}%
where $\hat{s}_{j}$ denotes the spin-1/2 operator at the $j\,$th site, $L$
denotes the total number of sites. The driving parameter $\lambda $
represents the ratio between the NNN coupling and the nearest-neighbor (NN)
coupling. The GS properties of the model (\ref{Ham}) has been widely studied
by both analytical method \cite{Haldane,Giamarchi} and numerical method \cite%
{Okamoto,Castilla,RChitra,SRWhite96}. The QPT driven by $\lambda $
is well understood. The driving term due to $\lambda $ is irrelevant
when $\lambda <\lambda _{c}(\simeq 0.2411)$, and the system flows to
a spin fluid or Luttinger liquid with massless spinon excitations.
As $\lambda
>\lambda _{c}$, the frustration term is relevant and the ground
state flows to the dimerized phase with a spin gap open
\cite{Haldane,Giamarchi}. The transition from spin fluid to
dimerized phase is known to be of BKT type \cite{Haldane,Giamarchi},
for which the transition point was hard to be determined numerically
due to the problem of logarithmic correction \cite{Affleck}. The
critical value of $\lambda _{c}=0.2411\pm 0.0001$ has been
accurately determined by various numerical methods
\cite{Okamoto,Castilla,RChitra,SRWhite96}.
\begin{figure}[tbp]
\includegraphics[width=9cm]{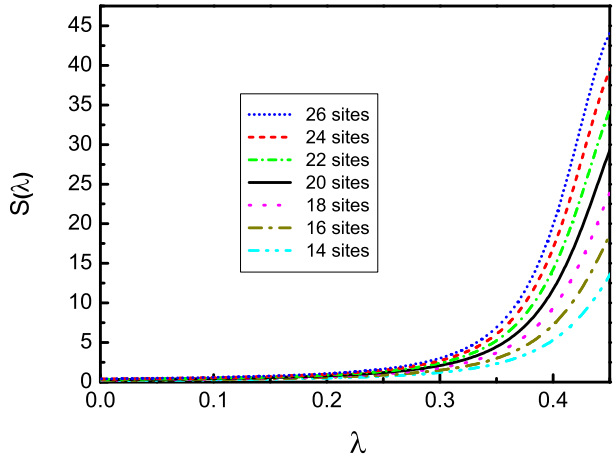}\newline
\caption{The GS fidelity susceptibility of the heisenberg chain
with next-nearest-neighbor interaction for finite system size from
14 sites to 26 sites. Obviously, there is no expected peaks can be
observed.} \label{Figure1}
\end{figure}

The GS fidelity for the model (\ref{Ham}) has been studied in Ref. \cite%
{schen07pre} and also in Ref. \cite{WangXG} in terms of operator
fidelity. No singularities in the GS fidelity or operator fidelity
around $\lambda_c$ have been detected for the system with different
sizes, which implies that the GS fidelity may be not an effective
characterization of the BKT-type QPT in this model. The BKT-type QPT
is a infinite order phase transition where the $n$-th order
derivative of GS energy is continuous.
\begin{figure}[tbp]
\includegraphics[width=9cm]{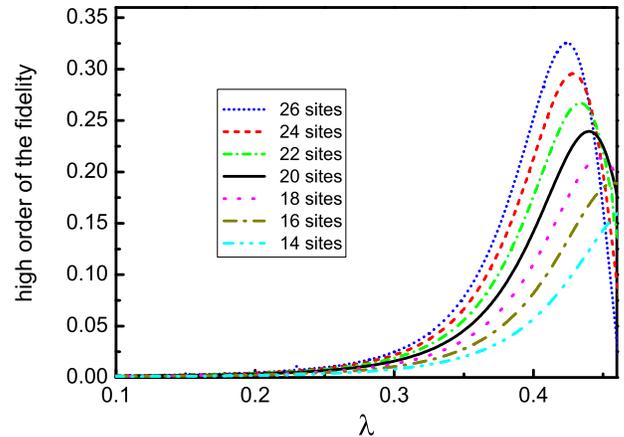}\newline
\caption{The third order term of the GS fidelity of the spin chain
with next-nearest-neighbor interaction for the finite system size
from 14 sites to 26 sites. Explicit peaks can be observed in this
figure. As the system size increases, the position of the peak gets
closer to the BKT-transition point.} \label{Figure2}
\end{figure}
\begin{figure}[tbp]
\includegraphics[width=9cm]{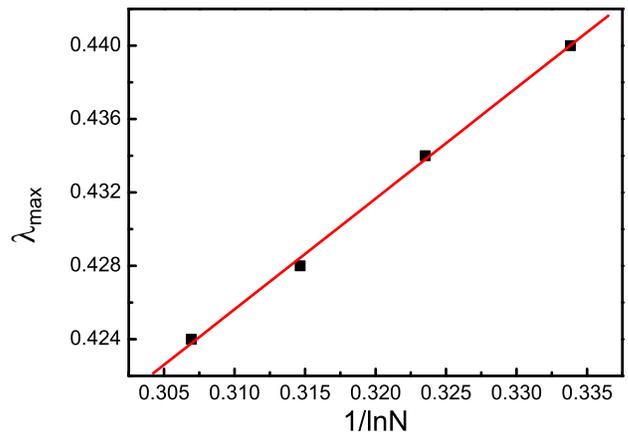}\newline
\caption{Finite-size scaling of the extrema of the third term of the GS
fidelity. A linear fit is made. According to this fit, when it comes to the
point $N\rightarrow \infty$, $\protect\lambda_c=0.238 \pm0.006$. }
\label{Figure3}
\end{figure}
In light of the higher-order fidelity being more powerful than its energy
judgement, we study the possibility for detecting the infinite-order
BKT-type QPT via the 3rd order fidelity and focus on the QPT in the spin
chain with NNN interactions as a concrete example. We first calculate the GS
wave functions by using the numerical exact diagonalization method for
finite size system, and thus the fidelity susceptibility and the 3rd order
fidelity can be extracted from the overlap of neighboring GS wave functions.
In Fig.1, we display the fidelity susceptibility for systems with different
sizes. We observe that no an obvious peak for the fidelity susceptibility is
detected in a wide range of the parameter $0<\lambda<0.5$. This result
suggests that the transition point for the BKT-type QPTs cannot be very
effectively characterized by the fidelity susceptibility either for a
finite-size system.

The BKT-type phase transition generally is an infinite order phase
transition for which the infinite order derivatives of the ground-state
energy is continuous. A good example with exact proof is the BKT-type
transition happened in the antiferromagnetic XXZ spin chain model \cite%
{YangCN}. In the BKT-type transition point, it has been proven analytically
that all the $n$-th order derivatives of ground state energy is continuous
\cite{YangCN}. Since the $n$-th order fidelity is much divergent than its
correspondence of derivative of the ground-state energy, one might expect
that there exists the possibility that the $n$-th order fidelity is
divergent even its $n$-th order energy derivative is continuous.
To see whether a higher order fidelity works better than fidelity
susceptibility in detecting the BKT-type QPT happened in this model,
we calculated the 3rd order fidelity versus the driving parameter as
shown in Fig. 2. It is clear that a peak is developed in the 3rd
order fidelity and
the location of peaks tends to get close to the side of transition point $%
\lambda _{c}$ with the increase of lattice size. To extrapolate the
$\lambda _{c}$ in the infinite size limit, we analyze the finite
size scaling of position of peak in the Fig. 3. When the system size
comes to infinity, the extrapolated value of the phase transition
point is $\lambda _{c}=0.238\pm 0.006$, which, within the scope of
fitting error, agrees well with $\lambda
_{c}=0.2411\pm 0.0001$ obtained by highly accurate numerical methods \cite%
{Okamoto,Castilla,RChitra,SRWhite96}.

\section{Summary}

\label{sec:sum}

We have shown the formulism for the high order of the fidelity in detail and
applied it to a concrete model, \textit{i.e.}, the one dimensional
Heisenberg chain with NNN interaction. We first calculate the ground-state
wavefunction of the system by exact diagnolization method, and then extract
fidelity susceptibility and the third order of the GS fidelity. We find that
despite the GS fidelity and the fidelity susceptibility being not a very
effective detector, the BKT-type phase transition happened in this spin
chain model might be effectively detected by the 3rd order term of the GS
fidelity for finite-size system. Although the physical meaning of the higher
order term of the GS fidelity hasn't been deeply understood, we wish that
our observation would stimulate further studies on this issue.

\begin{acknowledgments}
This work is supported by NSF of China under Grant No. 10821403,
programs of Chinese Academy of Sciences, National Program for Basic
Research of MOST, China and the Earmarked Grant Research from the
Research Grants Council of HKSAR, China (Project No. CUHK 400807).
\end{acknowledgments}

\end{document}